\documentclass{jfm}
\usepackage{graphicx}
\usepackage{epstopdf, epsfig}
\usepackage{xcolor}

\shorttitle{Guidelines for authors}
\shortauthor{D. Greenblatt, H. Müller-Vahl and C. Strangfeld}

\title{Airfoil Boundary Layer Bubble Separation and Transition in a Surging Stream}

\author{David Greenblatt\aff{1} \corresp{\email{davidg@technion.ac.il}},
  Hanns Müller-Vahl\aff{1}
  \and Christoph Strangfeld\aff{2,3}}

\affiliation{\aff{1} Faculty of Mechanical Engineering, Technion - Israel Institute of Technology, 3200003 Haifa, Israel
\aff{2}Department of Non-Destructive Testing, Bundesanstalt für Materialforschung und -pr{\"u}fung, Unter den Eichen 87, 12205 Berlin, Germany
\aff{3}Hermann-F{\"o}ttinger-Institut, Technische Universit{\"a}t Berlin, 10623 Berlin, Germany}

\begin{document}

\maketitle

\begin{abstract}
The effect of high-amplitude harmonic surging on airfoil laminar separation bubbles was investigated theoretically, and experimentally in a dedicated surging-flow wind tunnel. A generalized pressure coefficient was developed that accounts for local static pressure variations due to surging. This generalization, never previously implemented, facilitated direct comparisons between surging and quasi-steady pressure coefficients, and thus unsteady effects could be distinguished from Reynolds number effects. A momentum-integral boundary layer analysis was implemented to determine movement of the bubble separation point, and movement of the transition point was extracted from experimental surface pressure coefficients. The most significant finding was that bubble bursting occurs, counterintuitively, during early imposition of the favorable temporal pressure gradient. This is because the favorable pressure gradient rapidly drives the bubble aft, rendering it unable to reattach. Bursting resulted in large lift and form-drag coefficient oscillations, and failure to implement the generalized pressure coefficient definition resulted in temporal form-drag errors of up to 400 counts. \\
\end{abstract}


\section{Introduction}

The unsteady response of boundary layers resulting from the periodic pitching of airfoils has received widespread attention \citep{mccroskey1982unsteady,carr1988progress,corke2015dynamic}, particularly due to the importance of dynamic stall. In contrast, boundary layer behavior due to surging of the stream or airfoil has received far less attention. This is, in part,  because the generation of large amplitude surging oscillations relative to the mean velocity $(\sigma \equiv \Delta U/\bar{U})$ at typical flight or turbine Reynolds numbers  $(Re \equiv \bar{U}c/\nu)$ poses significant technical challenges. Early experiments on surging airfoils were performed by \cite{retelle1978unsteady} and \cite{brendel1988boundary}. More recently, studies that include surging have been performed by \cite{granlund2014airfoil}, \cite{choi2015surging}, \cite{dunne2015dynamic}, \cite{medina2018highamplitude}, and \cite{kirk2019vortex}. These investigations focus on the phenomenon of dynamic stall and are limited to low average Reynolds numbers $Re \leq 10^5$.

Recently, two independent relatively high $\{Re,\sigma\}$ experiments under low angle-of-attack pre-stall conditions \citep{strangfeld2016airfoil, zhu2020unsteady} were performed on NACA 0018 airfoils with nominally attached boundary layers, namely $\{3\times10^5, 0.5\}$ and $\{1.5 \times 10^6, 0.2\}$. In the former, at dimensionless surging frequencies of $k_{su} \equiv \omega c/2 \bar{U}=0.1$, comparisons of the lift coefficient as a function of phase angle, $C_l(\phi)$, to classical thin airfoil theory \citep{greenberg1947airfoil,isaacs1945airfoil, vanderwall1994on} were contradictory. With a tripped boundary layer, the lift coefficient corresponded qualitatively with theory, but a high-frequency oscillation was observed following deceleration of the stream. Without boundary layer tripping, the data produced an opposite effect to the theory, again with a high-frequency oscillation that was assumed to be associated with the shedding of a laminar separation bubble. In the later investigation, significant deviations from theory were observed, and when $k_{su}$ was increased from 0.025 to 0.05, high frequency oscillations were evident during the deceleration phase. Time-resolved background-oriented Schlieren images indicated that both curvature and position of the trailing-edge streaklines vary within the phase, indicative of a moving stagnation point that violates the theoretical Kutta condition.

The significant deviations from theory described above served as motivation to investigate temporal pressure gradient effects on laminar separation bubbles. The surface pressure coefficients, that identify the bubble separation point and transition, were used for this purpose. To do this, a generalized pressure coefficient was defined, which must be employed to account for temporal pressure gradient effects. This critically important generalization has never been implemented previously. To isolate purely unsteady effects from the influence of Reynolds number, pressure coefficients associated with surging were compared with those generated under quasi-steady conditions. In addition, a momentum-integral approach was adopted to predict movement of the bubble separation point. Finally, the effect of surging on the integral aerodynamic coefficients was evaluated.

\section{Theoretical Considerations}
\subsection{Pressure Coefficient under a Temporal Pressure Gradient}
\label{press-grad}
In streamwise surging flows, the test article pressure coefficient must be defined relative to the local freestream static pressure $p_{st} = p_{st}(x,t)$, which is not independent of $x$. In general, $p_{st}$ varies due to varying ambient conditions, such as exit-flow blockage, and under the assumption of incompressibility, namely $U=U(t)$, the pressure gradient $\partial p_{st}/\partial x$ depends on the freestream acceleration $dU/dt$. Notably, $p_{st}$ cannot be unambiguously measured along the tunnel walls because, in general, wall pressures respond to both test section pressure variations and test article load variations. To avoid dealing with this inherent coupling, $p_{st}$ is determined in the following manner. With the symmetric airfoil at $\alpha=0$, the local leading-edge $(x/c = 0)$ static pressure is determined by the simultaneous measurement of the stagnation pressure and freestream velocity, namely ${p_0}(0,t)$ and $U(t)$. Under the assumption that the leading-edge stagnation pressure coefficient $C_{p}(0,t) \equiv [{p_0}(0,t) - {p_{st}}(0,t)]/{q}(t)=1$, the corresponding static pressure at $x=0$ is calculated according to:
\begin{eqnarray}
{p_{st}}(0,t) = {p_0}(0,t) - {q}(t),
\label{le_stag_press}
\end{eqnarray}

\noindent where ${q}(t) = {\raise0.5ex\hbox{$\scriptstyle 1$} \kern-0.1em/\kern-0.15em \lower0.25ex\hbox{$\scriptstyle 2$}}\rho U^2$. To determine the local static pressure along the chord of the test article, Euler’s equation in the freestream is written as follows:
\begin{eqnarray}
\frac{1}{\rho}\frac{{\partial {p_{st}}}}{{\partial x}} = - \frac{{d U}}{{dt}}
\label{euler}
\end{eqnarray}

During experiments, $U(t)$ is measured, but for purposes of illustration we assume that:
\begin{eqnarray}
U(t) = \bar{U}(1 + \sigma \sin \omega t)\
\label{usfs}
\end{eqnarray}

\noindent where $\bar{U}$ is the mean freestream velocity. Substituting eqn. (\ref{usfs}) into eqn. (\ref{euler}) results in:
\begin{eqnarray}
\frac{1}{\rho}\frac{{\partial {p_{st}}}}{{\partial x}} = - \bar{U}\sigma \omega \cos \omega t
\label{pgrad}
\end{eqnarray}

\noindent which can be integrated as follows:
\begin{eqnarray}
\int_{p_{st}(0,t)}^{p_{st}(x,t)} {d{p_{st}}}  =  - \int_0^x {\rho {\bar{U}}\sigma \omega \cos \omega t dx}
\end{eqnarray}

\noindent to produce: 
\begin{eqnarray}
p_{st}(x,t) = p_{st}(0,t) - x\rho \bar{U}\sigma \omega \cos \omega t
\label{static_Press(t)}
\end{eqnarray}

Finally, for an arbitrary test article surface pressure $p(x,t)$, the local pressure coefficient is defined as: 
\begin{eqnarray}
c_{p}(x,t) \equiv [p(x,t) - p_{st}(x,t)]/{q}(t)
\end{eqnarray}

\noindent which, after substitution of eqn. (\ref{static_Press(t)}), results in:
\begin{eqnarray}
c_{p}(x,t) = [p(x,t) - p_{st}(0,t) + x\rho \bar{U}\sigma \omega \cos \omega t ]/q (t)
\label{cp_final}
\end{eqnarray}

\noindent or
\begin{eqnarray}
c_{p}(x,t) = \frac{p(x,t) - {p_{st}}(0,t)} {q(t)} + C(x,t)
\label{correct-cp}
\end{eqnarray}

\noindent where the first term on the right-hand side is the standard, or uncorrected, pressure coefficient $c_{pu}$, and correction term:

\begin{eqnarray}
C(x,t) = 4 \sigma k_{su} \hat{x} \frac{\cos \omega t}{(1+\sigma \sin \omega t)^2},
\label{CorrectionC}
\end{eqnarray}

\noindent with $\hat{x} \equiv x/c$. In contrast to pitching airfoils, where quasi-steady effects are assumed for $k_{pi} \equiv \omega c /2 U <0.002$ (e.g. Wickens, 1985), quasi-steady surging flows ($c_{pu} \approx c_{p}$) require that $C(c,t) \ll 1$ for all $t$.

\subsection{Unsteady Integral Boundary Layer Analysis}
\label{quasi-steady}
Unsteady effects on an airfoil boundary layer, and prediction of the bubble separation point, can be achieved via relatively simple application of von K\'{a}rm\'{a}n integral equation. Consider an airfoil boundary layer subjected to an external boundary condition, or edge-velocity, $U_e(s,t)$, where the coordinate $s$ is measured from the stagnation point, along the surface of the airfoil. The unsteady integral equation has the form:
\begin{eqnarray}
\frac{{{\tau _w}}}{\rho} = \frac{\partial ({U_{e}^{2}}\theta)}{{\partial s}} + {\delta^*}U_e\frac{{\partial U_e}}{{\partial s}} + \frac{\partial ({U_e \delta^*})}{{\partial t}}
\label{usvonKarman}
\end{eqnarray}

\noindent which, apart from the time-dependence expressed in the last term, is identical to the standard steady form (e.g. Schlichting, 1979). Equation \ref{usvonKarman} presents a ``closure'' problem, because a presently-unknown theoretical or empirical expression for $\delta^*(t)$ must be introduced (Docken, 1982). Here we circumvent this problem by using the quasi-steady version of eqn. \ref{usvonKarman} which we justify by neglecting the temporal pressure gradient term, expanded below, compared to the spatial one, namely
\begin{eqnarray}
\left | \delta^* \frac{\partial U_e}{{\partial t}} + U_e \frac{\partial \delta^*}{\partial t} \right | \ll \left | {\delta^*}U_e\frac{{\partial U_e}}{{\partial s}} \right |
\label{pg-ineq}
\end{eqnarray}

This assumption is based on two factors. First, for commonly encountered unsteady conditions, apart from a small fraction of the airfoil chord where $\partial c_p / \partial \hat{s} \approx 0$ (the suction peak), the first term is smaller than the third term (see below), namely:
\begin{eqnarray}
\left | 4 \sigma k_{su}  \frac{\cos \omega t}{(1+\sigma \sin \omega t)^2} \right | < \left | \frac{1}{\sqrt{1-c_p}} \frac{\partial c_p}{\partial \hat{s}} \right |
\label{pgnd-ineq}
\end{eqnarray}

\noindent where $c_p = 1 - \hat{U}_{e}^{2}$ from the mechanical energy equation, $\hat{U}_{e} \equiv U_e(s,t)/U(t)$ and $\hat{s} \equiv s/c$. This inequality is valid for the present investigation. Second, the first and second terms in eqn. \ref{pg-ineq} have opposite signs and tend to cancel each other out. The quasi-steady assumption allows us the considerable simplification of employing the well-known Pohlhausen velocity profile (Schlichting, 1979), together with the generalized boundary layer parameter (Docken, 1982):
\begin{eqnarray}
K \equiv \frac{\theta^2}{\nu} \left( \frac{{\partial U_e}}{{\partial s}} + \frac{1}{U_e}\frac{{\partial U_e}}{{\partial t}} \right )
\label{Pohl_param}
\end{eqnarray}

\noindent where $K = - 0.1567$ indicates the separation point. Thus the flow state, and in particular boundary layer separation, is determined by a combination of the spatial adverse pressure gradient produced by the airfoil geometry and angle of attack:
\begin{eqnarray}
K_{s} = \frac{\theta^2}{\nu} \frac{\partial U_e} {\partial s} = \hat{Q} \frac{\partial \hat{U}_e} {\partial \hat{s}}
\label{K-x}
\end{eqnarray}

\noindent and the  temporal pressure gradient produced by the surging stream:
\begin{eqnarray}
K_{t} = \frac{\theta^2}{\nu U_e} \frac{\partial U_e}{\partial t} = 2 \hat{Q} \sigma k_{su} \frac{\cos \omega t}{(1+\sigma \sin \omega t)^2}
\label{K-t}
\end{eqnarray}

\noindent where the dimensionless quadrature term is:
\begin{eqnarray}
\hat{Q} = \hat{Q}(\hat{s},t) =  \frac{0.47}{\hat{U}_{e}^{6}} \int_{0}^{\hat{s}} \hat{U}_{e}^{5} d \hat{s}
\label{Quad}
\end{eqnarray}

In order to compute the time-dependent separation point, we use the vortex-lattice method described in Drela (1989) to obtain $\hat{U}_e$ and $\partial \hat{U}_e / \partial \hat{s}$, for a given $\alpha$. The former facilitates the quadrature calculation of eqn. (\ref{Quad}), and hence $K_s$ and $K_t$ are computed. The expression $K_s + K_t = -0.1567$ is then solved numerically as a function of the phase-angle $\phi \equiv \omega t  $ to obtain $\hat{s}_\textnormal{\scriptsize{sep}}$ and hence $\hat{x}_\textnormal{\scriptsize{sep}}$. Note that the \emph{quasi-steady flow conditions}, corresponding to $K_t=0$ and referred to in section \ref{press-grad} are different to the the \emph{quasi-steady boundary layer assumptions} invoked in this section. 

\begin{figure}
\centering
\includegraphics[width=4in, trim={0 3.5cm 0 3.5cm},clip]{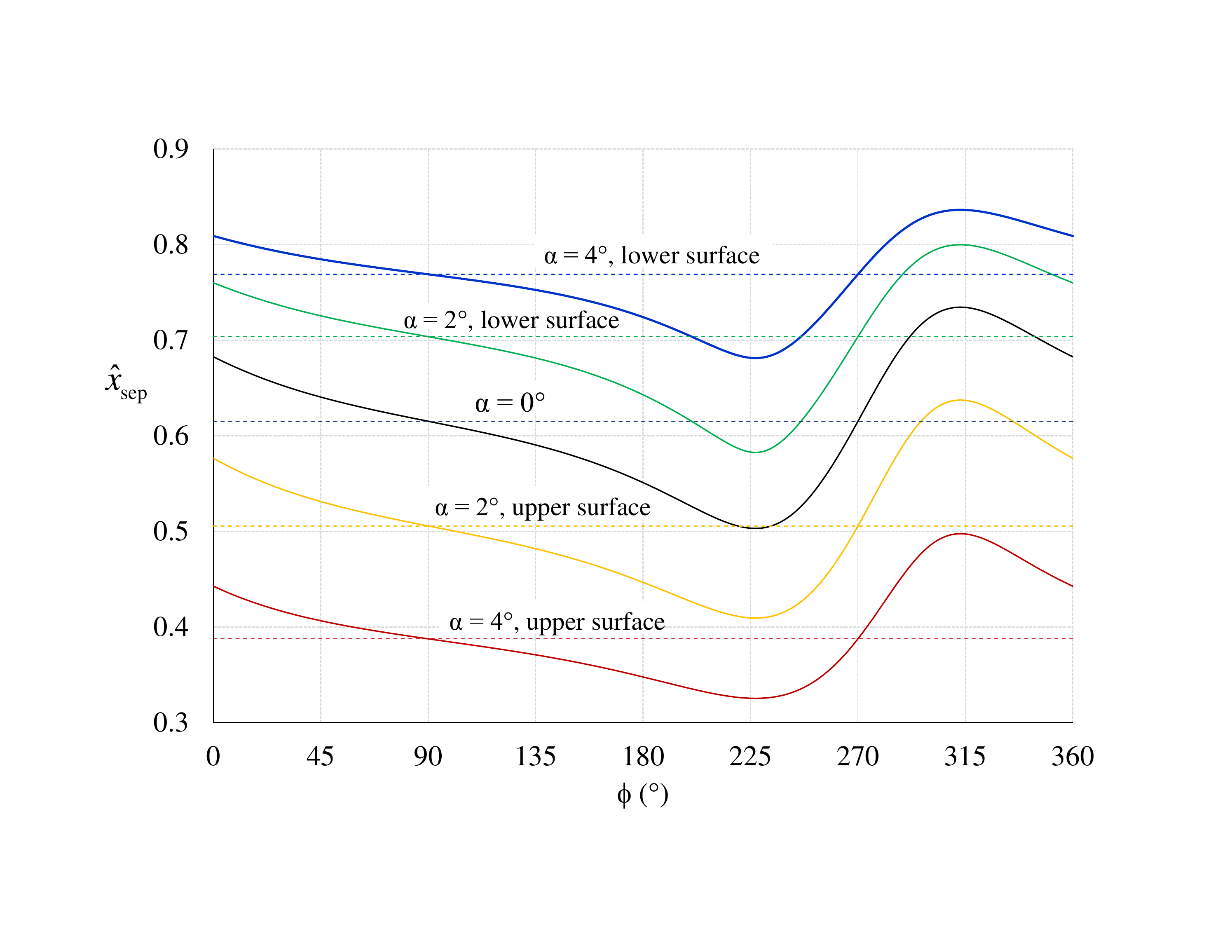}
\caption{Dimensionless chordwise separation point as a function of phase angle based on quasi-steady boundary layer assumptions; solid lines: quasi-steady momentum integral equation; dashed lines: quasi-steady flow corresponding to $K_t=0$.}
\label{us-separation}
\end{figure}

The separation point variation as a function of phase-angle is shown for three angles of attack, together with instantaneous static values ($K_t = 0$) in Figure \ref{us-separation}. As expected, the acceleration and deceleration phases are associated with downstream and upstream movement of the separation separation point respectively. However, from $\phi = 228 ^{\circ}$ to $314 ^{\circ}$, there is a relatively rapid downstream movement of the separation point. We will show in the sections below, that this plays a decisive role in the bursting of the separation bubble.  

\section{Experimental Setup \& Methods}
\label{sec:exp_setup}
Experiments were carried out at the Technion's Unsteady blow-down Low-Speed Wind Tunnel, that incorporates a 1004 mm $\times$ 610 mm test section and was configured with a 4.07 m test section length (see \citeauthor{greenblatt2016unsteady}, \citeyear{greenblatt2016unsteady}). A louver system, mounted at the test section exit and driven by a servo-motor, was used to vary the exit area and hence the freestream velocity. A NACA 0018 (chord $c=348$ mm and span of $b=610$ mm) was mounted 0.91 m downstream of the test section nozzle. It was equipped with a 1.2 mm wide passive slot at 5\% chord, which acted as a boundary layer trip. The tripped (slotted) side of the airfoil was defined as the upper surface, i.e., the low pressure surface for $\alpha > 0^{\circ}$, and the smooth side as the lower surface.  Forty four  symmetrically disposed pressure ports were close coupled to piezoceramic pressure transducers inside the model. Six additional pressure ports were located at chordwise positions of $\hat{x}=21.5\%$ and $69.5\%$ on the suction surface as well as $\hat{x}=69.5\%$ on the pressure surface at a distance of 100 mm from each side wall. Surface pressures were continuously recorded at a sample rate of 499 Hz and subsequently phase-averaged, while simultaneously, the simple average of two hot-wire measurements ahead of the airfoil (above and below) was used to determine $U(t)$. Harmonic surging and quasi-steady data were acquired between $\alpha = -4^{\circ}$ and $4^{\circ}$ in steps of $1^{\circ}$ and at $8^{\circ}$, under the conditions $k_{su} =  0.1$, $Re=3\times10^5$ and $\sigma=0.5$. Full details of the experimental setup are provided by \cite{strangfeld2016airfoil}.

\section{Results \& Discussion}
\begin{figure}
\centering
\includegraphics[width=4in]{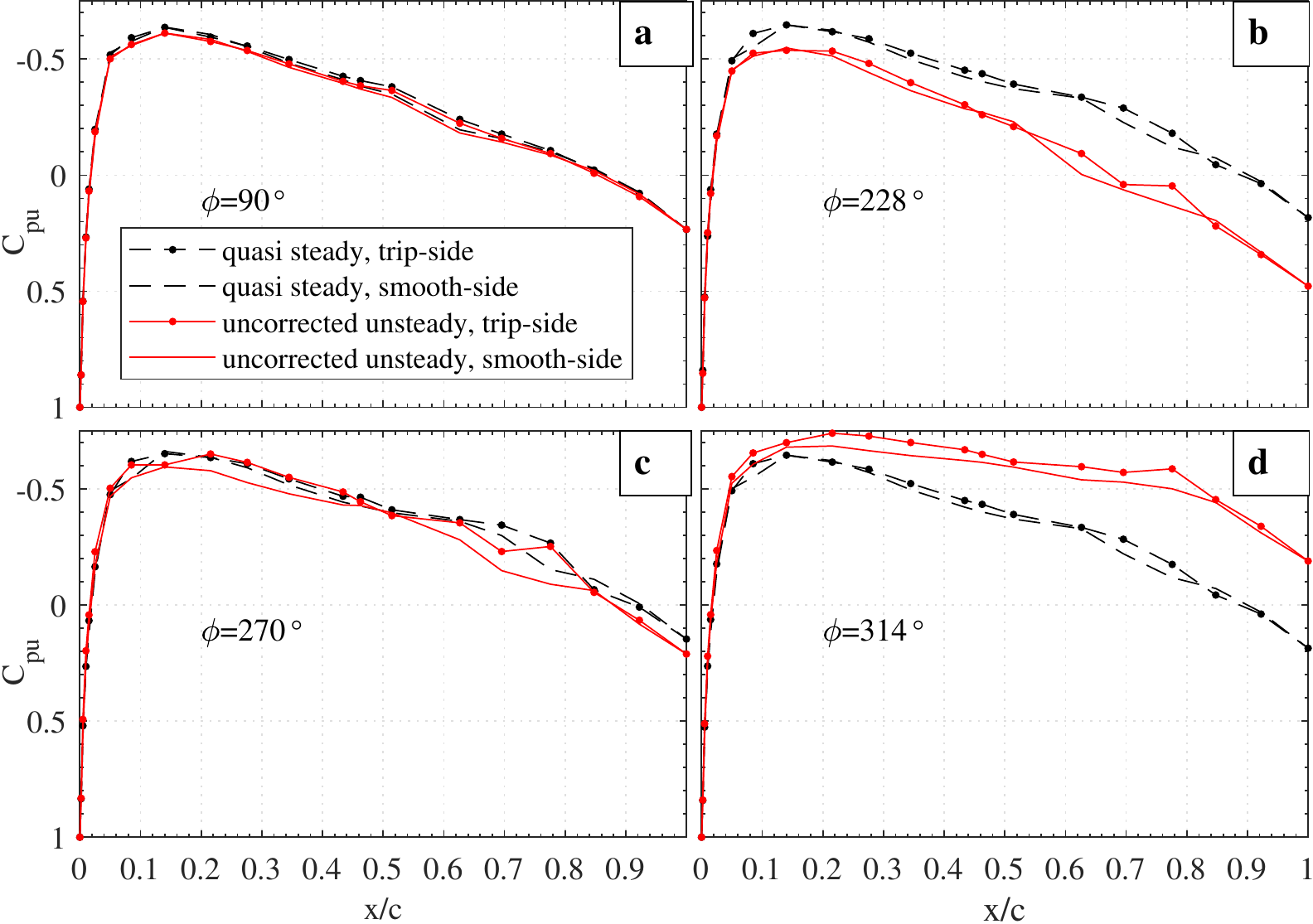}
\caption{Comparison of uncorrected unsteady and quasi-steady pressure coefficients at phases corresponding to $C=0$ (a, c) and  $\partial C/ \partial \phi = 0$ (b,d) for $\alpha = 0^\circ$.}
\label{uncorrected-cp}
\end{figure}

The results presented in this section rely on a direct comparison between unsteady and quasi-steady pressure coefficient distributions at identical Reynolds numbers, in order to infer the effects of unsteadiness on the boundary layer. Quasi-steady experimental pressure coefficients were generated by interpolating the data generated at eleven wind speeds, according to the method described in  \cite{muellervahl2020dynamic}. To illustrate the importance of the static pressure correction described in section \ref{press-grad}, consider the uncorrected unsteady pressure coefficient data, referenced to the static pressure at the leading-edge, compared to the quasi-steady data at $\phi=90 ^{\circ}$, $228^{\circ}$, $270^{\circ}$ and $314^{\circ}$ in Figure \ref{uncorrected-cp}. The first and third phase angles correspond to $C=0$, while the second and fourth angles correspond to the correction minimum and maximum, namely $\partial C/ \partial \phi = 0$, respectively. For $C=0$, valid comparisons between unsteady and quasi-steady pressure coefficients are possible because the streamwise pressure gradient is zero. Therefore, the unsteady pressure gradient effects on the boundary layer can be inferred by comparing the unsteady and quasi-steady data. In contrast, when the streamwise pressure gradient is non-zero as exemplified by the $\phi = 228^{\circ}$ and $314^{\circ}$ angles, the effect of the pressure gradient on the boundary layer cannot be inferred because the data are biased by the incorrect static pressure. The bias increases linearly with $\hat{x}$ as shown in eqn. (\ref{CorrectionC}). Consequently, effects \emph{of the pressure gradient on the separation bubble} cannot be separated from the effects \emph{of the pressure gradient itself}. Thus comparisons at all phase angles presented below are based on the generalized pressure coefficient expressed in eqns. (\ref{correct-cp}) and (\ref{CorrectionC}), where the expression for $U(t)$ in eqn. (\ref{usfs}) is replaced with the hot wire measurements described in section \ref{sec:exp_setup}.

\begin{figure}
\centering
\includegraphics[width=5in]{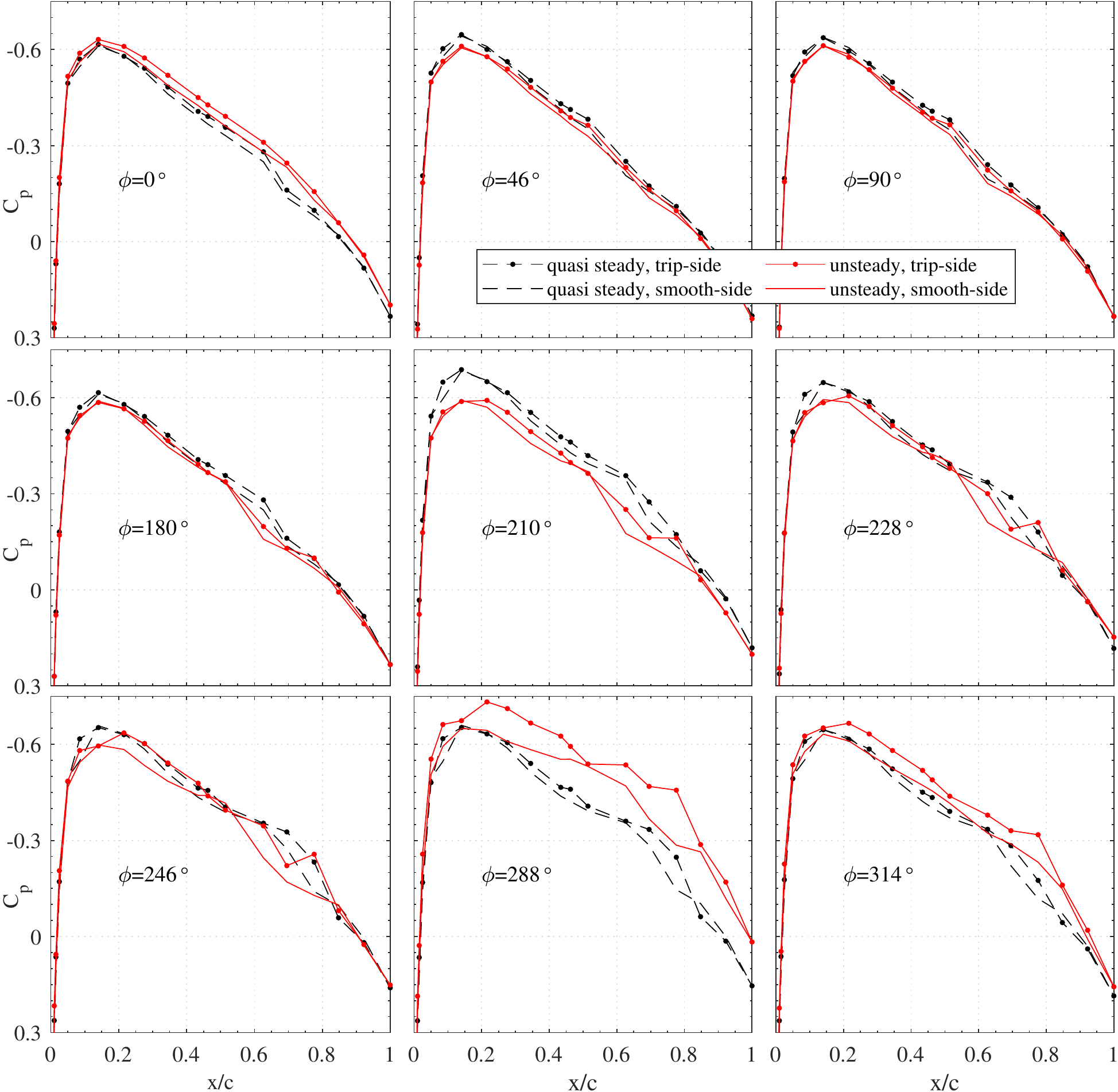}
\caption{Corrected unsteady and quasi-steady pressure coefficients at selected representative phase angles for $\alpha = 0^\circ$.}
\label{Cp-comp}
\end{figure}

Surging and quasi-static pressure coefficients are compared at $\alpha=0^{\circ}$ at selected representative phase angles in Figure \ref{Cp-comp}. High spatial resolution NACA 0018 pressure measurements by \cite{kurelek2016coherent} show that the upstream location of almost constant pressure indicates bubble separation ($\hat{x}_\textnormal{\scriptsize{sep}}$), a sharp pressure rise indicates transition ($\hat{x}_\textnormal{\scriptsize{tran}}$) and an abrupt decrease in slope indicates reattachment ($\hat{x}_\textnormal{\scriptsize{att}}$). From our relatively low spatial resolution data, $\hat{x}_\textnormal{\scriptsize{sep}}$ is difficult to pinpoint, but the transition location $\hat{x}_\textnormal{\scriptsize{tran}}$ is identifiable. In the middle of the acceleration phase ($\phi=0^{\circ}$) the quasi-steady data show bubbles with transition at $\hat{x}=0.63$ for both the trip-side and the smooth-side. In contrast, for the surging case there is no clear evidence of bubble transition on the trip-side and a downstream movement of transition with a weaker pressure rise on the smooth-side. If we assume that changes in $\hat{x}_\textnormal{\scriptsize{sep}}$ and $\hat{x}_\textnormal{\scriptsize{tran}}$ correlate, then we can infer that the separation point moves downstream, which is consistent with the theoretical considerations presented in section \ref{quasi-steady}. 

As the acceleration diminishes, bubble transition in both the surging and quasi-steady cases moves gradually upstream and at the peak velocity ($\phi=90^{\circ}$), without any temporal pressure gradient, the transition locations are indistinguishable at $\hat{x}=0.51$. We can thus conclude that history effects are small to negligible at this phase. As the stream decelerates, the main observable difference is that the quasi-steady transition points move downstream, purely on account of the reducing Reynolds number (quasi-steady data at $\phi=0^{\circ}$ and $180^{\circ}$ are identical). Thus the net effect of the deceleration is to move the bubble upstream. Again, assuming that changes in $\hat{x}_\textnormal{\scriptsize{sep}}$ and $\hat{x}_\textnormal{\scriptsize{tran}}$ correlate, the difference between the surging and quasi-steady data is consistent with the theoretical considerations of section \ref{quasi-steady}. During the second part of the deceleration phase, beginning at $\phi=192^{\circ}$, there is a curious pressure drop between $\hat{x}=0.69$ and $0.78$ on the trip-side (Figure \ref{Cp-comp}), that increases in relative terms and is still observed at $\phi=270^{\circ}$.  This may be indicative of a second turbulent  bubble or a bubble structure that is not typical of that observed in steady flows.

\begin{figure}
\centering
\includegraphics[width=4in,trim={0 3.5cm 0 0},clip]{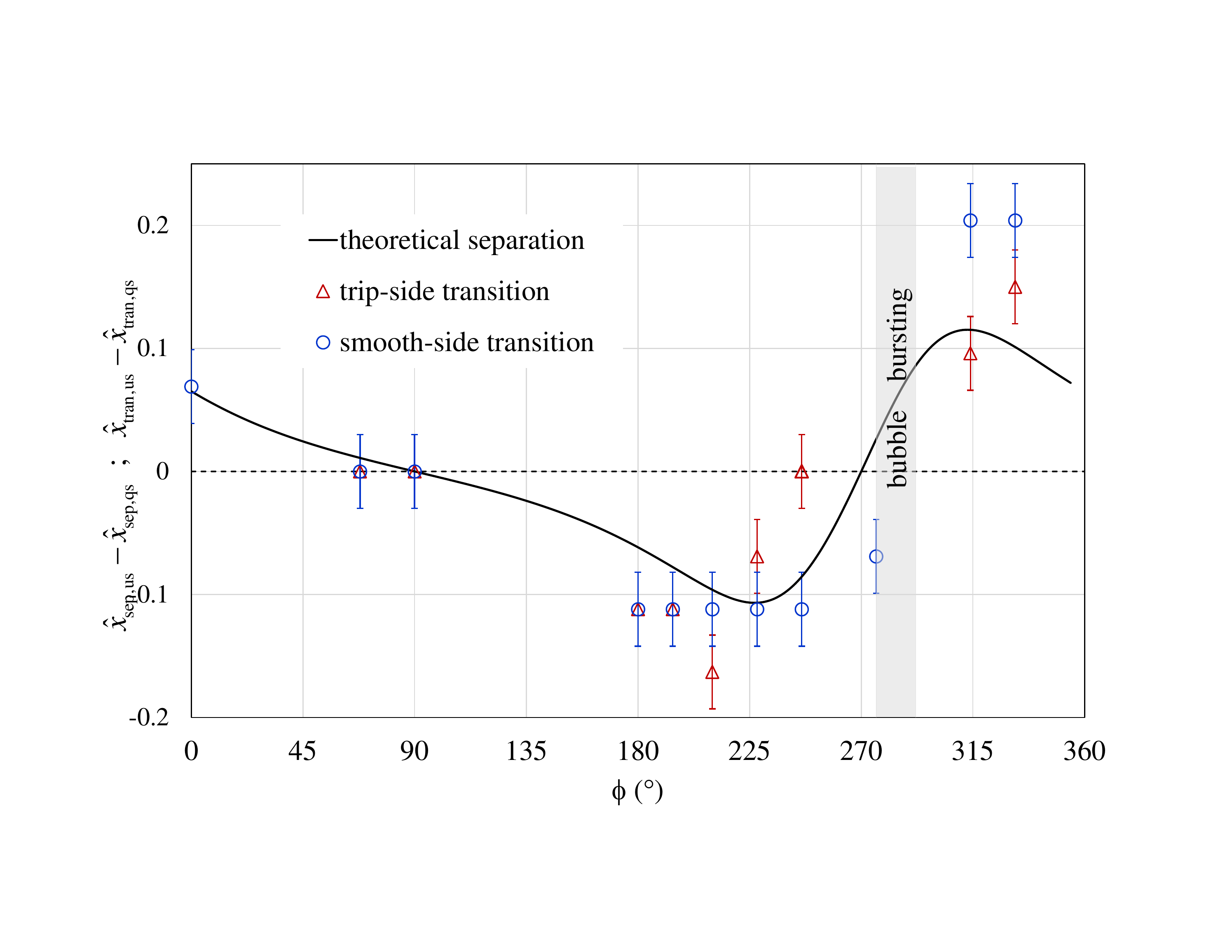}
\caption{Movement of the bubble separation point according to the momentum integral analysis and movement of the transition points gleaned from experimental data at $\alpha=0^{\circ}$.}
\label{SepTranComp}
\end{figure}

From the angle $\phi=276^{\circ}$ to $\phi=288^{\circ}$ there is a rapid and significant drop in pressures on the aft part of the airfoil (Figure \ref{Cp-comp}). This is a clear indication that either one or both of the bubbles do not reattach, resulting in ``bursting'' or ``venting.'' The assumed correlation between the movement of the separation and transition points is quantified by comparing the theoretical difference $\hat{x}_{\textnormal{\scriptsize{sep}},\textnormal{\scriptsize{us}}} - \hat{x}_{\textnormal{\scriptsize{sep}},\textnormal{\scriptsize{qs}}}$ (section \ref{quasi-steady}) with the experimental difference $\hat{x}_{\textnormal{\scriptsize{tran}},\textnormal{\scriptsize{us}}} - \hat{x}_{\textnormal{\scriptsize{tran}},\textnormal{\scriptsize{qs}}}$ in Figure \ref{SepTranComp}. Despite the relatively large uncertainty associated with determination of the transition location, the qualitative correlation is clear. Both the theoretical result and the experimental data reveal the mechanism of bubble bursting due to the effects of the temporal pressure gradient. Theoretically, bubble separation moves rapidly downstream between $\phi = 228^{\circ}$ and $314^{\circ}$, but bursting is only evident for $\phi > 270^{\circ}$ (see Figure \ref{SepTranComp}). Thus, in contrast to spatial adverse pressure gradients that give rise to bubble bursting, \emph{it is in fact during the favorable pressure gradient phase of the cycle that the bubble(s) burst(s).} The rapid downstream movement of the separation point precipitates full separation of the bubble. To the best of our knowledge this bubble bursting mechanism has not previously been observed in surging airfoil flows.

\begin{figure}
\centering
\includegraphics[width=2.5in]{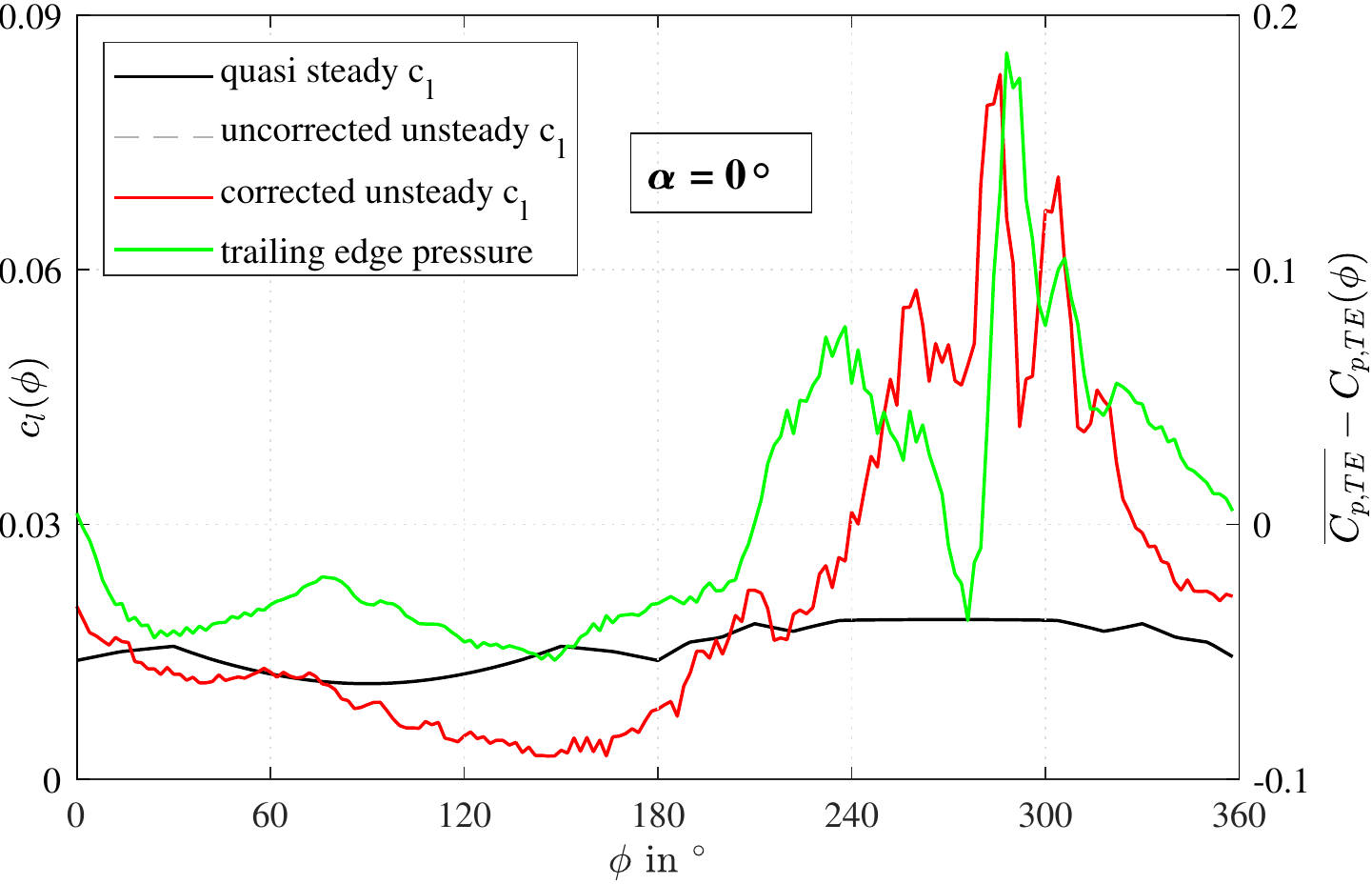}
\includegraphics[width=2.5in]{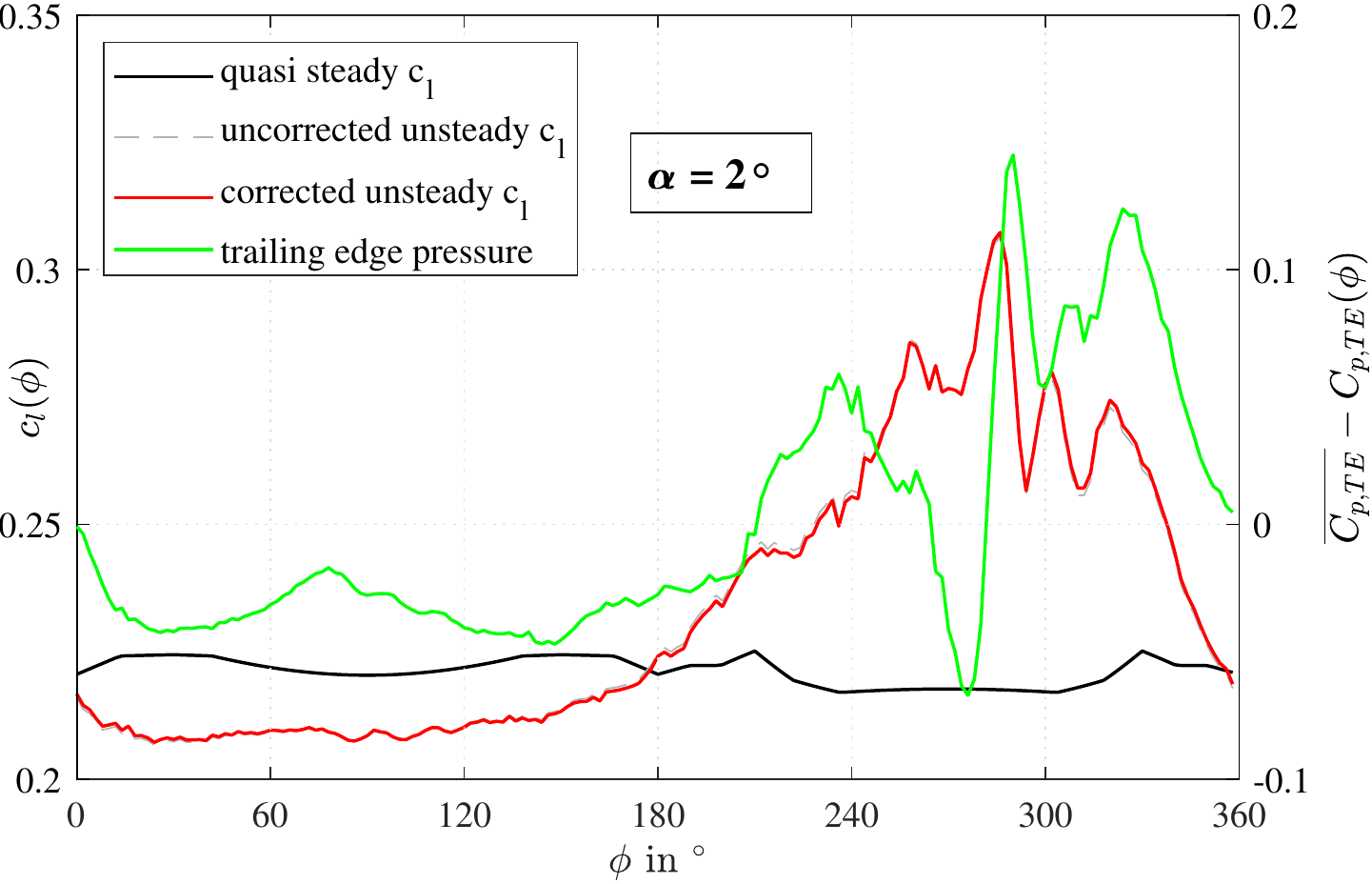}
\caption{Comparison of unsteady surging (corrected and uncorrected) and quasi-steady lift coefficients, as well as trailing-edge pressure variations.}
\label{cl-comp}
\end{figure}

\begin{figure}
\centering
\includegraphics[width=2.5in]{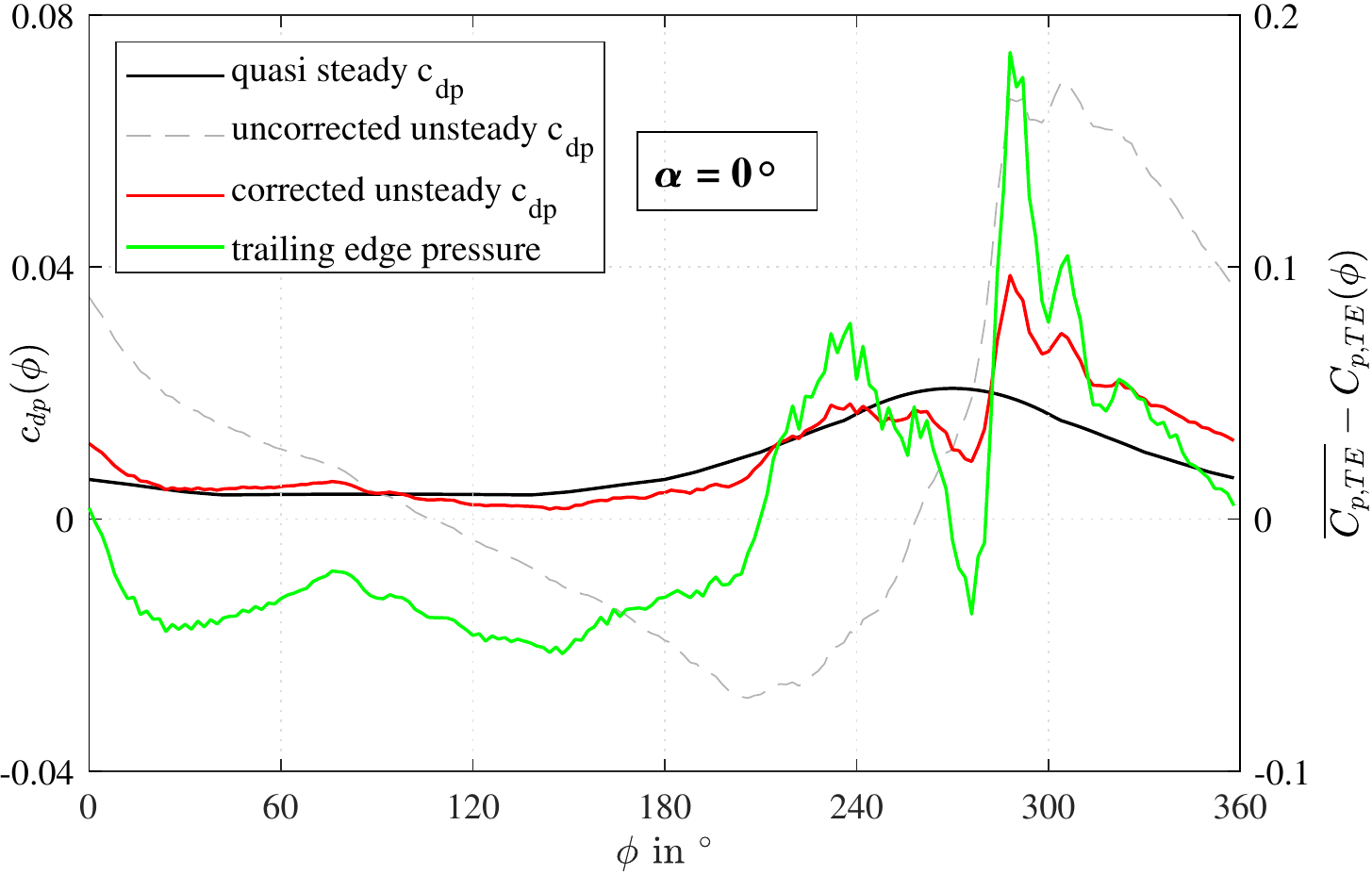}
\includegraphics[width=2.5in]{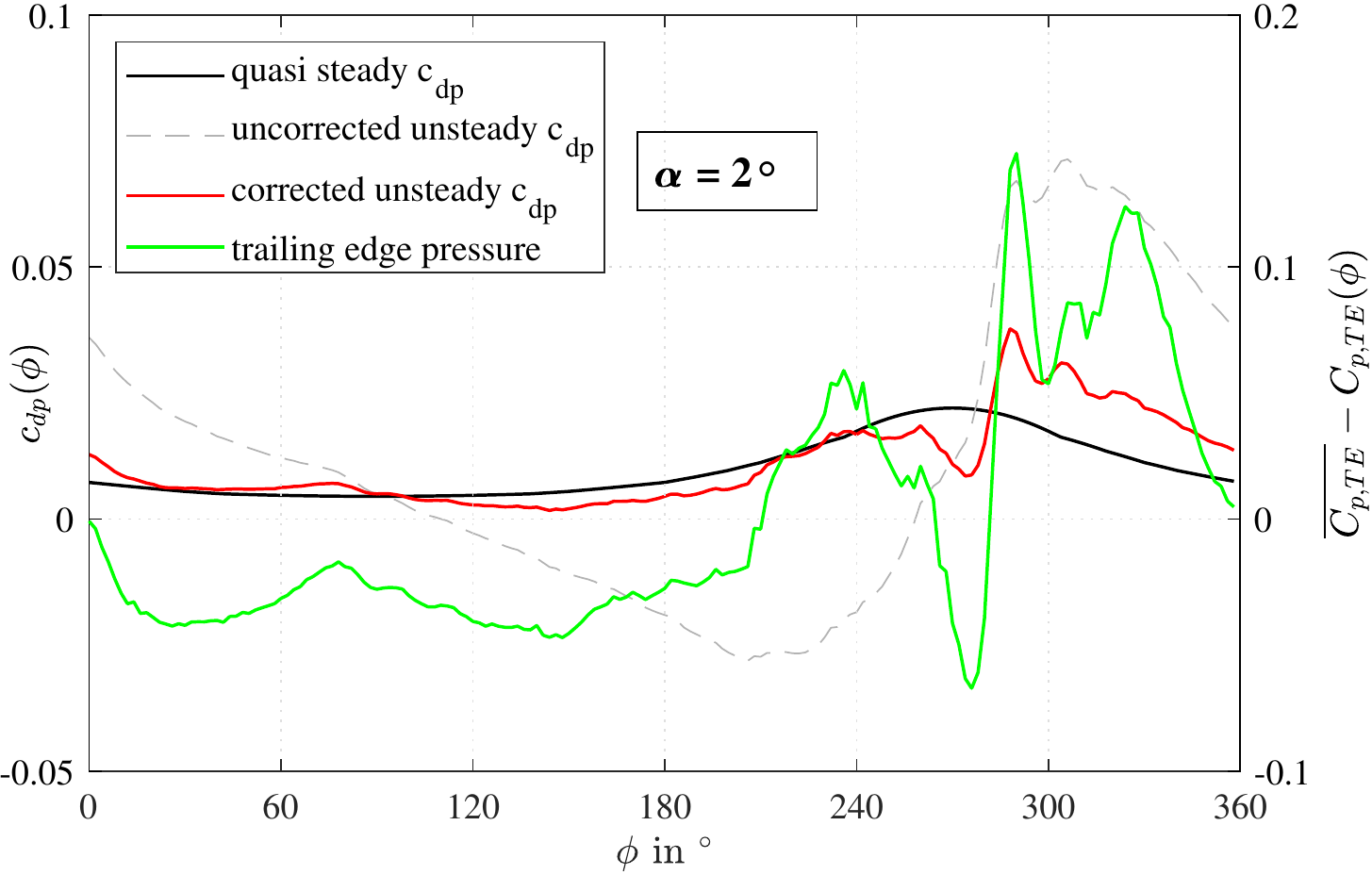}
\caption{Comparison of unsteady surging (corrected and uncorrected) and quasi-steady form-drag coefficients, as well as trailing-edge pressure variations.}
\label{cd-comp}
\end{figure}

Aerodynamic coefficients were determined by integrating the pressure coefficient results, employing the method described by Anderson (2011). The lift and pitching moment coefficients $c_l$ and $c_m$ are identical at $\alpha=0^{\circ}$ and virtually identical at $\alpha=2^{\circ}$ because both sides of the airfoil are affected equally by the correction term $C$ in eqn. \ref{correct-cp}. However, the form-drag $c_{dp}$ clearly is affected because the effects on both sides of the airfoil are additive. To illustrate this, both the corrected and uncorrected $c_l$ and $c_{dp}$ results are shown for $\alpha=0^{\circ}$ and $2^{\circ}$ in Figures \ref{cl-comp} and \ref{cd-comp} respectively, together with the quasi-steady results and trailing-edge pressure coefficient. The lift coefficient results at both angles of attack are qualitatively similar and representative of the $0^{\circ} \leq\alpha \leq 4^{\circ}$ range (not shown); this indicates a similar stalling mechanism at these positive angles. The trailing-edge pressure correlates with the integrated coefficients in the sense that local pressure minima coincide with $c_l$ changes ($\partial c_l / \partial \phi$). The $c_l$ increase starting at around $\phi=200^{\circ}$ appears to be associated with the local pressure drop between $\hat{x}=0.69$ and $0.78$ discussed above, while the peaks at $\phi=288^{\circ}$ and $\phi=306^{\circ}$ are associated with bubble bursting. Similar qualitative results for $0^{\circ} \leq\alpha \leq 4^{\circ}$ suggest, in fact, that for $\alpha > 0^{\circ}$ the smooth-side lower surface bubble bursts prior to that on the tripped-side. This is because the smooth-side favorable spatial pressure gradient moves bubble separation further downstream (Figure \ref{us-separation}) and thus imposition of the favorable temporal pressure gradient moves the bubble too far downstream to reattach, prior to that of the trip-side. Hence, the second peak, which has a smaller relative value at higher $\alpha$ is associated with bursting of the trip-side bubble.
\\
The effect of the pressure coefficient correction on $c_{dp}$ (Figure \ref{cd-comp}) is dramatic with differences of up to 400 form-drag counts ($\textnormal{form-drag count} = 10^{4} c_{dp}$). Indeed, failure to implement the generalized pressure coefficient results in large negative and positive, physically unrealistic, values. As expected, and in contrast to $c_l$ (Figures \ref{cl-comp}), $c_{dp}$ and trailing-edge pressure peaks correlate directly. On the one hand, the corrected unsteady versus quasi-steady comparison allows us to distinguish between purely Reynolds number effects and combined Reynolds number and unsteady effects, and on the other hand, Figure \ref{SepTranComp} allows us to interpret the results. Specifically, the local $c_{dp}$ reduction between $\phi=260^{\circ}$ and $270^{\circ}$ is due to upstream movement of the separation and transition points that produce higher trailing-edge pressures. This precedes the relatively large $c_{dp}$ peak that occurs due to lower surface bubble bursting, precipitated by rapid downstream movement of the bubble separation point. Differences between phase-averaged unsteady and quasi-steady $c_{l}$ and $c_{m}$ are small; $c_{l}$ increases and $c_{m}$ decreases due to surging, are both $\mathcal{O}(10^{-2})$. The phase-averaged form-drag coefficient, increases by up to 75 drag counts  between $-2^{\circ} \leq\alpha \leq 4^{\circ}$. However, there is a crossover at $\alpha \approx -2.5^{\circ}$, where surging reduces $c_{dp}$ by 40 and 50 counts at $\alpha = -3^{\circ}$ and $-4^{\circ}$, respectively.


\section{Conclusions}

The effect of unsteady, periodic, surging flows on airfoil separation bubble behavior at low angles of attack was evaluated theoretically and experimentally. A generalized pressure coefficient was defined, for the first time, and the effects of surging were determined by comparing unsteady and quasi-steady pressure coefficients. Furthermore, a quasi-steady momentum integral boundary layer analysis was employed to predict changes in the bubble separation point. The analysis and data pointed to a stabilizing effect during the latter part of the acceleration that produced downstream movement of the bubble. Deceleration, on the other hand, produced upstream movement of the bubble, when Reynolds number effects were accounted for. Counterintuitively, bubble bursting occurred during the early part of the acceleration phase. This was precipitated by rapid downstream movement of bubble separation, shown theoretically and experimentally, accompanied by large lift and form-drag coefficient oscillations. Lift and form-drag coefficient behavior were qualitatively similar for  $0^{\circ} \leq\alpha \leq 4^{\circ}$ which suggests that the smooth, lower surface, bubble bursts prior to that on the tripped upper surface. In future research, detailed high spatial resolution pressure and flowfield measurements will be made to precisely quantify the complex flowfield produced by temporal pressure gradients.

\bibliographystyle{jfm}
\bibliography{NDT_library.bib}

\end{document}